\documentclass[twocolumn,pra,superscriptaddress,nofootinbib]{revtex4}

\usepackage{graphicx}
\usepackage{changes}
\usepackage{amssymb}
\usepackage{amscd}

\usepackage{tikz}
\usetikzlibrary{chains}
\usetikzlibrary{fit}
\usepackage{pgflibraryarrows}		
\usepackage{pgflibrarysnakes}		
\usepackage{xcolor}
\usepackage{epsfig}
\usepackage{amsthm}
\usepackage{amsmath}
\usepackage{amsfonts}
\usepackage{amssymb,amstext}
\usepackage[colorlinks=true,urlcolor=blue, hyperindex] {hyperref}
\usepackage{epigraph}

\usepackage{etoolbox}

\makeatletter
\patchcmd{\epigraph}{\@epitext{#1}}{\itshape\@epitext{#1}}{}{}
\makeatother

\theoremstyle{plain}

\theoremstyle{definition}

\definecolor{nblue}{rgb}{0.2,0.2,0.8}
\definecolor{ngreen}{rgb}{0.2,0.7,0.2}
\definecolor{nred}{rgb}{0.8,0.2,0.2}
\definecolor{nblack}{rgb}{0,0,0}

\definecolor{darkgreen}{rgb}{0,0.4,0}
\definecolor{orange}{rgb}{1,0.5,0}

\begin{document}

\title{Was inflation necessary for the existence of time?}
\author{Sandra Stupar}
\affiliation
{Institute for Theoretical Physics, ETH Z\"urich, 8093 Z\"urich, Switzerland}
\author{Vlatko Vedral}
\affiliation
{Clarendon Laboratory, University of Oxford, Parks Road, Oxford OX1 3PU, United Kingdom}
\affiliation{Centre for Quantum Technologies, National University of Singapore, 3 Science Drive 2, Singapore 117543} 
\affiliation{Department of Physics, National University of Singapore, 2 Science Drive 3, Singapore 117542}

\begin{abstract}
Modern physics has unlocked a number of mysteries regarding the early Universe, such as the baryogenesis, the unification of the strong and electroweak forces and the nucleosynthesis. However, understanding the very early Universe, close to the Planck epoch, still presents a major challenge. The theory of inflation, which is assumed to have taken place towards the end of the very early Universe, has been introduced in order to solve a number of cosmological problems. However, concrete observational evidence for inflation is still outstanding and the physical mechanisms behind inflation remain mostly unknown.
In this paper we argue for inflation from a different standpoint. In order for time to have any concrete physical meaning in the very early and the early Universe, the capacity of the Universe to measure time - its size or, equivalently, memory - must be at least as large as the number of clock ``ticks'' that need to be recorded somewhere within the Universe. Using this simple criterion, we provide a sketch proof showing that in the absence of inflation the subsystems of the Universe might not have been able to undertake the synchronised evolution described by the time we use today.
\end{abstract}

\maketitle

The theory of inflation came out of the need to describe the large scale features of the Universe as we observe them today. In contrast to the theory of Big Bang, inflation offers us relatively simple explanations for the {\it flatness problem} (the fact that the cosmological constant today has a value extremely close to 1 --- $\Omega \approx 1.012$), the {\it horizon problem} (distant regions of the Universe should have been at the causal contact in the past), the {\it monopole problem} (Big Bang theory without inflation predicts too many magnetic monopoles to have been created - albeit these have not been observed) and others (the interested reader is encouraged to have a look at~\cite{Guth, GuthEternalInflation}). Another argument in support of inflation comes from the assumption that the observable and the unobservable Universe might be entangled~\cite{VlatkoEntanglement}. 

We will not restrict ourselves to a specific inflationary scenario (such as {\it e.g. }eternal inflation), but will consider general theory that assumes exponential expansion in the very early Universe, usually taken to have happened between around $10^{-37}s$ and $10^{-32}s$~\cite{Guth, Linde, AlbrechtSteinhardt}. The most important variable in this theory is the number of so called $e$-folds the Universe had to go through during inflation, {\it i.e.} the factor of exponential expansion that we will label $C$, and which we will lower bound in our approach.
 
Our argument provides additional explanation in support of the theory of inflation - namely, the existence of time and its passage as we record it. We term this the {\it `universal clock problem'}. Our results tell us that the inflation at the beginning of the Universe was maybe needed for the notion of time to be physically meaningful. They also lead us to propose an intriguing starting point of the Universe: the Universe comes into being at the point when its smallest size is large enough to record its own temporal evolution. Our calculations are based on the well-established quantum metrological and general relativistic bounds, linking the entropy and energy of the Universe to its capacity to store and process information.

Physical quantum clocks designed` to keep track of time need to produce ticks, that can be described in the form of the so-called {\it tick registers} (see {\it e.g.}~\cite{ClockModel}). The Universe itself also needs to be able to store this information about the ticks/time it provides. The ticks produced should be stored in distinguishable quantum states. 
We will therefore make the following assumption: 
{\it In order for the common notion of time to be possible, the Universe itself has to contain (represent) a universal clock with a large enough memory that can store its own `ticks'. In particular, the Universe needs to have at least as many possible distinguishable states as ticks it has produced as a clock.}

We can also relate our approach to more specific thermodynamical quantum clocks, where it was shown that the accuracy and ability of the clock to store time directly relates to its increase in entropy~\cite{ThermodynamicQClock}. In our case, the increase in the entropy of the Universe as a thermodynamical clock simply means the increase in the number of possible distinguishable states. This allows the Universe to keep counting ticks and measuring universal time. We will show how the ability to store time information is directly related to the speed of the evolution of the Universe which then relates to its entropy increase via the Bekenstein formula.  

Our approach fits well within the Page-Wootters picture~\cite{PageWootters}. A number of authors~\cite{PageWootters,Wootters, Banks, Mott, Brout, Briggs} have shown that one can define `time evolution' of a system via its correlations with the rest of the Universe that acts as a clock. One can hence start from the Wheeler-DeWitt equation~\cite{DeWitt}, which states that the Universe as a closed system is stationary.  An important assumption that has to be made is that the clock system is independent from the rest of the Universe whose `time' it attempts to measure ({\it i.e.} there is no interaction term in the total Hamiltonian). Some of the obstacles in this approach have recently been overcome~\cite{GiovannettiQuantumTime, ChiaraVlatko}. Even though we will not use the Page-Wootters picture explicitely, one can consider a cosmological clock that is represented by the size (radius) or some other specific variable of the Universe, which is to a good degree decoupled from the observers. 

Note that we are not speculating about the fundamental origins of what time we experience really is or how best to measure it. These aspects belong to another fundamental area of research. Our argument here is simply that for an experience of time we use in physics to be possible (be it an illusion, a comfortable assumption or a fundamental entity) our Universe needed to go through an exponential expansion at the very beginning.

Below we will present calculations for the case with and without inflation, first taking the Planck time as a fundamental ticking interval of the Universe as a clock. In our second approach we will rely on the Margolus-Levitin bound to obtain minimal time interval needed for a Universe to change to an orthogonal state and hence for a tick to happen. We will see that in both approaches we obtain similar lower bounds on the factor of exponential expansion of the Universe at the beginning. We note that our calculations are based on the FLRW model of the Universe~\cite{ReviewFLRW}.\\

{\it Bekenstein bound and the number of ticks of the universal clock} 
\\

It is known that the entropy of a black hole is the maximum entropy that can exist within a certain area, hence we can use the Hawking-Bekenstein bound to put an upper bound on the entropy of the Universe at the very beginning~\cite{BekensteinEntropy, HawkingRadiation, HawkingBlackHoles}. We use the spherically symmetric model of the Universe and denote the entropy of the black hole by $S_{BH}$, the entropy of the Universe $S_U$ and the smallest interval in which universal clock could tick by $\Delta t$. The following then holds:

\begin{equation}
\label{eq:HBbound}
2^{\frac{S_{BH}}{k_B}}\geq 2^{\frac{S_U}{k_B}}\geq\frac{t}{\Delta t } 
\end{equation}
which simply encapsulates the fact that the number of distinguishable states in the universe must be at least as large as the number of ticks $\frac{t}{\Delta t }$, where $t$ is the total time of evolution of the Universe. 
Since $S_{BH}=\frac{k_BA}{4 l_p^2}$ and $A(t)=4 \pi l(t)^2$, we have that:
\begin{equation}
2^{\frac{ \pi l(t)^2}{l_p^2}}\geq \frac{t}{\Delta t }
\label{eq:Bekenstein}
\end{equation}
where $l(t)$ is the radius of the Universe at time $t$ and $A(t)$ its area, $k_B$ Boltzmann constant and $2^{\frac{S_{BH}}{k_B}}$ is the maximum number of distinguishable states of the Universe. 
Note that Eq.~\ref{eq:Bekenstein} rests on the assumption that each tick of the Universe needs to be stored in a distinguishable state. Hence at each moment, the Universe needs to have at least as many distinguishable states as ticks it has produced.

We are mostly interested in the time after the Planck time $t_p\approx 10^{-44}s$ until soon after the end of the inflation. We will assume that inflation lasted until around $t_2=10^{-32}s$~\cite{Narlikar, Guth}. In the following calculations we will often omit the relevant units for simplicity.
We will proceed to show that, without inflation, the rate of the expansion would probably be too slow for the Universe to be able to create enough memory to keep track of its own time. 

Next we consider two scenarios, first with Planck unit of tick time and second using minimal orthogonal time approach to calculate the ticking interval.  \\

{\it Ticking in Planck units --- no inflation}
\\

Consider the expansion of the Universe after Planck's time $t_p$, without the inflation occurring. The following formula for the radius $l$ of the Universe at the later times should hold~\cite{Narlikar}:
\begin{align}
\label{eq:NoInf} 
l(t)=K\sqrt{t} \\
K=l^p/\sqrt{t_p}
\end{align}
where $l^p=l(t_p)$ is the radius of the Universe at Planck's time, and depends on the horizon of the Universe one is considering (e.g. particle horizon, event horizon, Hubble horizon,...). There is no uniquely agreed value and we will take $l^p=10^{-55}m$ as the radius of the observable Universe at the Planck time~\cite{Atkatz}. Hence $K\sim 10^{-33}$. Note that we use scaling with time of the cosmological scale factor to express the scaling of the radius of the Universe.

We will assume the Planck unit of time to be the smallest interval of time that is observable in the Universe, hence we set $\Delta t=t_p$. This stems from the accepted view that the times smaller than the Planck time are not detectable or even, possibly, not physically meaningful. In this scenario the Universe needs to `remember' and store its own state each Planck unit of time. 

From Eq.~\ref{eq:Bekenstein} it follows that:
\begin{align}
&2^{\frac{K^2\pi t}{l_p^2}} \geq \frac{t}{t_p}\\
&2^{10^{70}10^{-66}\pi  t} \geq 0.2*10^{44}t
\end{align}
The above inequality is not satisfied for approximately $10^{-44}<t<10^{-2}$. The Universe could not have expanded as given in \ref{eq:NoInf} at the beginning if we assume that it needs to be able to keep track of its own time. 

If we would, however, insert the current values of the size and age of the Universe, the above inequality would be satisfied. This inequality therefore describes the current Universe well, but for it to be able to keep track of time since the beginning it would need a more rapid evolution at the start.\\

{\it Planck units --- inflation}
\\

One of the implications of the previous analysis is that after the time we denote as $t_1$ (close to Planck's time), there had to be a different evolution equation for the size of the Universe, in order for it to be able to continue measuring its time. We therefore assume that there was an inflationary epoch ending around $t_2=10^{-32}s$~\cite{Narlikar} during which the Universe grew exponentially:

\begin{equation}
l_2=l(t_2)=l_1e^{C}
\end{equation}
where $C$ is the factor of expansion during the inflation and $l_1=l(t_1)$ radius of the Universe at the start of the inflation. We know $l_1>l^p\geq 10^{-55}m$ since at that moment Universe was at least as large as it was at Planck's time (usual start of the inflation is taken around $t_1=10^{-37}s$ or $10^{-36}s$~\cite{Narlikar, GuthPlanckData}). There is no consensus on the value of these parameters ($l_1$, $t_1$and $C$), nor is there any strong experimental evidence. 

Going back to Eq.~\ref{eq:Bekenstein}, we would like the following to hold:
\begin{equation}
\label{eq:HBInflation}
2^{\pi l_1^2e^{2C}/l_p^2}\geq\frac{t_2}{t_p}
\end{equation} 
Hence, we obtain an inequality between $C$ and $l_1$:
 \begin{align}  
 \label{eq:Expansion}                                                                                                                                                                                                                                                                                                                                                                       
 &l_1 \geq e^{-78 - C}
 \end{align}
 If {\it e.g.} the initial size of the Universe is $l_1=10^{-51}m$ the rate of expansion would be $C \geq 38$. If, on the other hand $l_1=10^{-55}m$ then $C \geq 48$.

We can combine above inequality in Eq.~\ref{eq:Expansion} with the fact that $l_1>l^p\geq 10^{-55}m$, and conclude that Eq.~\ref{eq:HBInflation} is certainly fulfilled for:
$10^{-55} \geq e^{-78 - C},\,\,C \geq 48$. Note that this conclusion is stronger than necessarily needed, but, if it holds, then our inequality~\ref{eq:Bekenstein} is satisfied and we know that Universe was able to still count its time at least until $t_2$. This, as previously shown, does not hold when using the model of expansion without inflation.

Result obtained agrees with most cosmological models that state that exponential factor should be around $C\approx 50$ or larger ($C$ number of $e$ - folds), depending on the problem the inflation is aimed to solve \cite{Narlikar, AlbrechtSteinhardt,GuthPlanckData} and the starting moment of the inflation. 
Our argument suggests that in order for inflation to solve the `universal clock' problem we need at least $48$ - fold inflation period to occur at the beginning of the Universe. \\

{\it Margolus-Levitin bound} \\

One should note that a similar approach to the one shown, can be used for a Universe that does not tick in Planck's units. For this approach we need to be able to approximate the average energy of the Universe at the time of our interest. Then one can obtain a minimal time needed for the Universe to change to a distinguishable (orthogonal) state using the Margolus-Levitin bound~\cite{MargolusLevitin} and follow the same reasoning as above. In this bound, unlike in the Heisenberg-Robertson uncertainty relation for time and energy~\cite{Heisenberg, Robertson}, one does not consider change in energy between two states, but uses the average energy instead.

The Margolus-Levitin bound is given by:

\begin{equation}
\label{eq:MLBound}
\Delta t \geq \frac{\pi \hbar}{2\overline{E}}
\end{equation}
where $\overline{E}$ is the average energy of the system and $\Delta t$ is the minimal time needed for a system to pass to an orthogonal (completely distinguishable) state.

We can bound the average energy by demanding that the radius of the Universe is not smaller than the Schwarzschild radius~\cite{Schwarzschild} of a system with the total energy of $E=Mc^2$:

\begin{align}
l\geq R_S=\frac{2GE}{c^4},\,\,
E\leq \frac{lc^4}{2G} \\
\Rightarrow \overline{E} \leq \frac{lc^4}{2G}
\end{align}
Inserting $l_p^2=\frac{\hbar G}{c^3}$ and using Eq.~\ref{eq:MLBound} we have:
\begin{equation}
\Delta t \geq \frac{\pi l_p^2}{lc},\,\,
l_p=c t_p
\end{equation}

Hence the following holds for the number of ticks:

\begin{equation}
\frac{t}{\Delta t} \leq \frac{lt}{\pi l_p t_p}
\end{equation}

To satisfy Eq.~\ref{eq:Bekenstein}, knowing that above inequality holds, we can assume:
\begin{equation}
\label{eq:MLBekenstein}
2^{\frac{\pi l^2}{l_p^2}}\geq \frac{lt}{\pi l_pt_p} \; .
\end{equation}

Note that, as in the chapter with the Planck unit of time, this condition is not necessary but is sufficient for Eq.~\ref{eq:Bekenstein} to hold and our Universe to create enough orthogonal states to store all of its ticks.

Let us label with $l(t)$ size of the Universe at time $t$, to obtain an inequality in terms of $y=\frac{l(t)}{l_p}$:
\begin{equation}
\label{eq:MLNoInflation}
\pi 2^{\pi y^2}\geq y t 10^{44}
\end{equation}
\\

{\it No inflation ---}
\\

 In the case of a non-inflationary model, we have that $l(t) \sim K \sqrt{t}$, where $K=l^p/\sqrt{t_p} \sim 10^{-33}$. Let us consider time $t=10^{-30}s$. Then $y= 10^{-13}$ and the inequality~\ref{eq:MLNoInflation} is not satisfied. This shows that, in this case, Universe could not have kept track of time until or after $10^{-30}s$. \\

{\it Inflation ---}
\\

Now consider scenario in which inflation takes place. We again take the end of the inflation at $t_2\approx10^{-32}s$, and labelling by $C$ the factor of exponential expansion: 
\begin{align}
\label{eq:exp}
l_2=l(t_2)=l(t_1)e^{C}
\end{align}
where $t_1$ denotes the beginning of the inflation. Again we now know that $l_1=l(t_1)>l^p=10^{-55}m$.

Imposing inequality~\ref{eq:MLNoInflation} to work for all $l_1$ we have
\begin{align}
&l_1e^C \geq 10^{-34}\\
&\Rightarrow C \geq 48
\end{align}

Depending on the exact value of $l_1$ smaller $C$ could be sufficient, but this bound works for any given $l_1>10^{-55}m$. Note that one would get a similar result with a bit different $C$ if we consider later times, $t>10^{-32}$. Hence our result is not dependent of taking this exact time $t_2$ for the end of inflationary period.

Interestingly, we have obtained the same bound as when considering Planck time as a unit time step of the universal clock. 

We have thus shown that our approach also works when using the Margolus-Levitin bound for the calculation of the smallest time interval of a tick of the universal clock. We see that the inflationary theory allows for the existence of time which the Universe has enough capacity to store. Our argument indicates that without the inflation, time and evolution as universal notions might not have existed.
\\

{\it Conclusions and future work}
\\

We have used few well-known laws and usual simplifying assumptions: the entropy of the Universe at any time is upper bounded by the one of the black hole of the same size, number of distinguishable states of a system is directly related to the entropy, Universe has a spherical symmetry and we use parametric time to express the tick interval of the universal clock. We do not claim the necessity of inflation for the existence of time, but our results show that the inflation with the factor of exponential expansion of around $50$ was sufficient for Universe to be able to keep track of its time in the early stages.

When considering the ticking of time, one naturally makes connections with the arrow of time. It is usually argued that the arrow of time arises from the $2^{nd}$ law of Thermodynamics, a law stating that the entropy of the closed system cannot decrease. As our Universe is considered to be a closed system (and may well be the only one such known to us), the $2^{nd}$ law should hold. Hence, the time flow that we notice follows the direction of the increasing entropy, and could not go `backwards'. 
Our argument provides another explanation for this, when viewed as a consequence of the fact that in order to time itself, the Universe needs to be able to provide more and more ticks, hence it needs to have more distinguishable states. From Eq~\ref{eq:HBbound} one concludes that the entropy of the Universe needs to increase (or at least not decrease) during the evolution. The flow of time again has one direction in this scenario, which is the direction of the increase in the number of ticks, {\it i.e.} the increase in the number of the possible states and with this the entropy of the Universe. 

Another way to think of the implications of the second law, is that the end of the Universe would come when it runs out of its clock time by reaching the maximum entropy. Hence the clock states of the Universe would be so mixed that there would be no more capacity to discriminate between different times. This notion is akin to the so called {\it heath death}, first postulated by Lord Kelvin. 

Our result is a novel stepping stone for the further investigation of the connection of inflationary theories and the notion of time and causality we experience. Theory of inflation can already explain some of the specific features of our Universe that makes it seem quite special, and we argue that existence of time might be another one of them. We hope that this work will encourage more research in this direction from the cosmological, quantum information and other physics communities. \\

{\it Acknowledgments}
\\

Authors thank Paul Davies and Volkher Scholz for insightful comments on the work. S. Stupar acknowledges support from the Swiss National Science Foundation (SNSF) via the National Centre of Competence in Research QSIT, as well as project No. $200020_{-}165843$. V. Vedral thanks the Oxford Martin School, the John Templeton Foundation and the EPSRC (UK). VV's research is also supported by the National Research Foundation, Prime Minister's Office, Singapore, under its Competitive Research Programme (CRP Award No. NRF- CRP14-2014-02) and administered by Centre for Quantum Technologies, National University of Singapore.

\bibliographystyle{unsrt}
\fontsize{3pt}{1pt}\selectfont
\bibliography{refInflation}

\end{document}